\documentclass[aps,prl,twocolumn,groupedaddress]{revtex4-1}
\usepackage[T1]{fontenc}       % Use modern font encodings

\usepackage{graphicx}  % needed for figures
\usepackage{amssymb}   % for math
\begin{document}

\title{Phonon-Plasmon Interaction in Metal-Insulator-Metal Localized Surface Plasmon Systems}

\author{Abdelali Mrabti}
\affiliation{Institut d'Electronique, de Micro-\'electronique et de Nanotechnologie (IEMN, CNRS-8520), Cit\'e Scientifique, Avenue Poincar\'e, 59652 Villeneuve d'Ascq, France}
\author{Ga\"etan L\'ev\^eque}
\affiliation{Institut d'Electronique, de Micro-\'electronique et de Nanotechnologie (IEMN, CNRS-8520), Cit\'e Scientifique, Avenue Poincar\'e, 59652 Villeneuve d'Ascq, France}
\email{gaetan.leveque@univ-lille1.fr}
\author{Rana Nicolas}
\affiliation{LIDYL, CEA, CNRS, Universit\'e Paris-Saclay, 91191 Gif-Sur-Yvette, France}
\author{Thomas Maurer}
\affiliation{Laboratory of Nanotechnology and Optical Instrumentation, UMR 6281 STMR, Technological University of Troyes, 12 Rue Marie Curie, CS 42060, 10004 Troyes Cedex, France}
\author{Pierre-Michel Adam}
\affiliation{Laboratory of Nanotechnology and Optical Instrumentation, UMR 6281 STMR, Technological University of Troyes, 12 Rue Marie Curie, CS 42060, 10004 Troyes Cedex, France}
\author{Abdellatif Akjouj}
\affiliation{Institut d'Electronique, de Micro-\'electronique et de Nanotechnologie (IEMN, CNRS-8520), Cit\'e Scientifique, Avenue Poincar\'e, 59652 Villeneuve d'Ascq, France}
\author{Yan Pennec}
\affiliation{Institut d'Electronique, de Micro-\'electronique et de Nanotechnologie (IEMN, CNRS-8520), Cit\'e Scientifique, Avenue Poincar\'e, 59652 Villeneuve d'Ascq, France}
\author{Bahram Djafari-Rouhani}
\affiliation{Institut d'Electronique, de Micro-\'electronique et de Nanotechnologie (IEMN, CNRS-8520), Cit\'e Scientifique, Avenue Poincar\'e, 59652 Villeneuve d'Ascq, France}

\begin{abstract}
	We investigate theoretically and numerically the coupling between elastic and localized surface plasmon modes in a system of gold nanocylinders separated from a thin gold film by a dielectric spacer of few nanometers thickness. That system supports plasmon modes confined in between the bottom of the nanocylinder and the top of the gold film, which arise from the formation of interference patterns by short-wavelength metal-insulator-metal propagating plasmon. First we present the plasmonic properties of the system though computer-simulated extinction spectra and field maps associated to the different optical modes. Next a simple analytical model is introduced, which allows to correctly reproduce the shape and wavelengths of the plasmon modes. This model is used to investigate the efficiency of the coupling between an elastic deformation and the plasmonic modes. In the last part of the paper, we present the full numerical simulations of the phononic properties of the system, and then compute the acousto-plasmonic coupling between the different plasmon modes and five acoustic modes of very different shape. The efficiency of the coupling is assessed first by evaluating the modulation of the resonance wavelength, which allows comparison with the analytical model, and finally in term of time-modulation of the transmission spectra on the full visible range, computed for realistic values of the deformation of the nanoparticle.
\end{abstract}

\maketitle

\section{Introduction}
Surface plasmons, which are optical modes supported by metal nanoparticles, have been thoroughly investigated for few decades due to their fascinating abilities to confine and enhance electromagnetic field in very sub-wavelength volumes \cite{schuller_plasmonics_2010}, at specific wavelengths ruled by the shape of the particle and its nearby environment. For those reasons, nanoscale-engineered systems have a large number of applications in domains like biosensing \cite{mock_local_2003,saison-francioso_plasmonic_2012,maurer_enhancing_2013}, enhanced-Raman spectroscopy \cite{wang_probing_2014}, metamaterials \cite{liu_metamaterials:_2011}, photo-thermal therapy \cite{jaque_nanoparticles_2014}, and plasmomechanics \cite{maurer_beginnings_2015}. Particularly interesting are systems composed of a small ensemble of closely-coupled metal particles, or of metal films interacting with one or several metal particles placed a few nanometers away \cite{maurer_coupling_2015,zhang_substrate-induced_2011}. In those regimes, the localized plasmons modes become very sensitive to a variation in length well below the nanometer. In case of extended contact area between a metal film and a metal particle (nanocubes, nanocylinders...), particular modes form where light is concentrated in between the two flat metal surfaces, with complex field distribution resulting from the interference of propagating plasmons constrained to move in between the particle and the film \cite{moreau_controlled-reflectance_2012,nicolas_plasmonic_2015}. The characteristic of the obtained modes will strongly depend on the thickness of the spacer layer and the shape of the cavity where the light is trapped. 

Besides, surface plasmons have been investigated since the middle of the 90s for their sensitivity to mechanical oscillations sustained by metal nanoparticles. The acousto-plasmonic interaction is generally investigated by pump-probe experiments \cite{voisin_time-resolved_2000,nelet_acoustic_2004}, where a wealth of phenomena are involved from the excitation of the metal particle by a laser pulse until it is back to equilibrium \cite{hodak_ultrafast_1998,hodak_size_1999,hartland_coherent_2002}, or by Raman spectroscopy \cite{bachelier_origin_2010,portales_probing_2008}. Beside their fundamental interest for the understanding of electronic and mechanical properties of metal nanosytems, these studies may have possible applications to mass sensing \cite{li_ultra-sensitive_2007,naik_towards_2009}.
If the coupling between surface plasmon modes and the mechanical oscillations sustained by metal nanosytems have been studied for a wide range of particle's shapes (spheres  \cite{van_dijk_detection_2005}, cubes \cite{staleva_vibrational_2008}, rods \cite{hu_vibrational_2003,zijlstra_acoustic_2008,soavi_ultrasensitive_2016,margueritat_surface_2006}, columns \cite{large_acousto-plasmonic_2009}, antenna \cite{della_picca_tailored_2016}, crosses \cite{obrien_ultrafast_2014}), investigations of system composed of interacting metal particle and metal film are still lacking, to the best of our knowledge.

The purpose of this paper is to investigate the acousto-plasmonic properties of a system made of an array of nanocylinders deposited on a substrate consisting in a thin (few nanometers thick) dielectric layer overcoating a gold film. It has been previously shown by us that such a coupled particle-substrate system supports so-called metal-insulator-metal localized-surface-plasmon modes where the light is essentially enhanced in between the cylinder and the metal film \cite{nicolas_plasmonic_2015}. We will see that the confinement of those modes in that portion of the spacer makes them very sensitive to deformation of that volume specifically. In the next section, we summarize the plasmonic properties of the system by presenting numerical extinction spectra together with electric field distributions. In the following section, we introduce a simple analytical model which allows to derive a closed-form expression giving the wavelengths of the plasmon modes as a function of the shape parameters of the portion of the spacer directly lying under the nanocylinder, and use that model to investigate the wavelength shift of plasmonic modes under different kind of deformation. In the final section, we present the full numerical simulations of the phononic properties of the same system and of the acousto-plasmonic interaction. We focus particularly on five elastic modes for which the elastic energy is essentially confined in or close to the nanocylinder. We show that a significant modulation of the transmission spectrum around the resonance wavelengths of the plasmons modes can be obtained, attributed to the particular sensitivity of the MIM-plasmon modes to the geometry of the cavity ; an application to pump-probe experiment is finally discussed.

\section{Plasmonic properties}\label{sect1}
\begin{figure}[h]
	\centering
	\includegraphics[width=8.5cm]{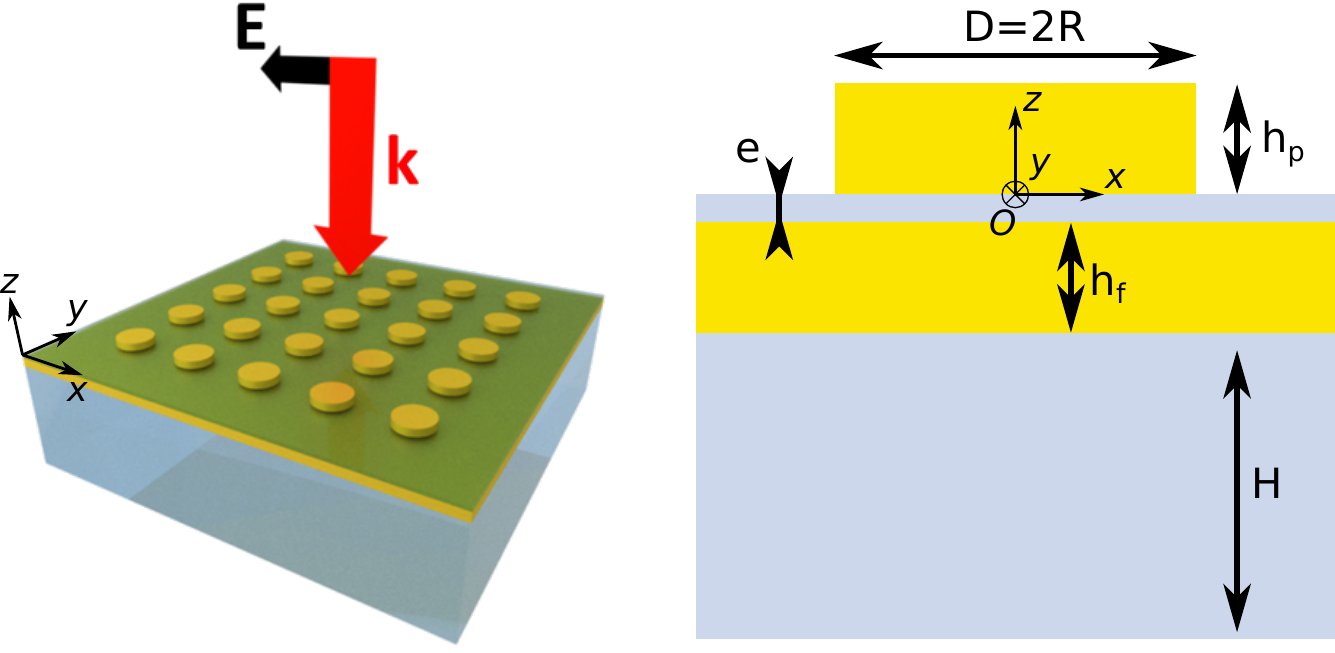}
	\caption{Geometry of the system and cross-section view indicating the shape parameters of the nanocylinder and of the membrane. The structure is optically excited by a plane wave in normal incidence on the AuNCs, linearly polarized along the $x$-axis. The point O is the center of the bottom face of the AuNC.}
	\label{fgr:fig1}
\end{figure}

The investigated system is presented in Fig.~\ref{fgr:fig1}, and consists in a square array of gold nanocylinders (AuNCs) deposited on a multilayered membrane composed of gold and silica. In the whole paper, the AuNCs have a radius $R=100$ nm and a height $h_p=50$ nm, the silica spacer is $e=6$ nm thick and has a refractive index of 1.5, the gold film is $h_f=50$ nm thick, and lies on a $H=144$ nm-thick membrane of silica. The period is $a=300$ nm. The refractive index data for gold are from Johnson and Christy \cite{johnson_optical_1972}. The grating is illuminated in normal incidence from the $z>0$ half space, the electric field being taken parallel to the $x$-axis.

\begin{figure}[h]
	\centering
	\includegraphics[width=8.5cm]{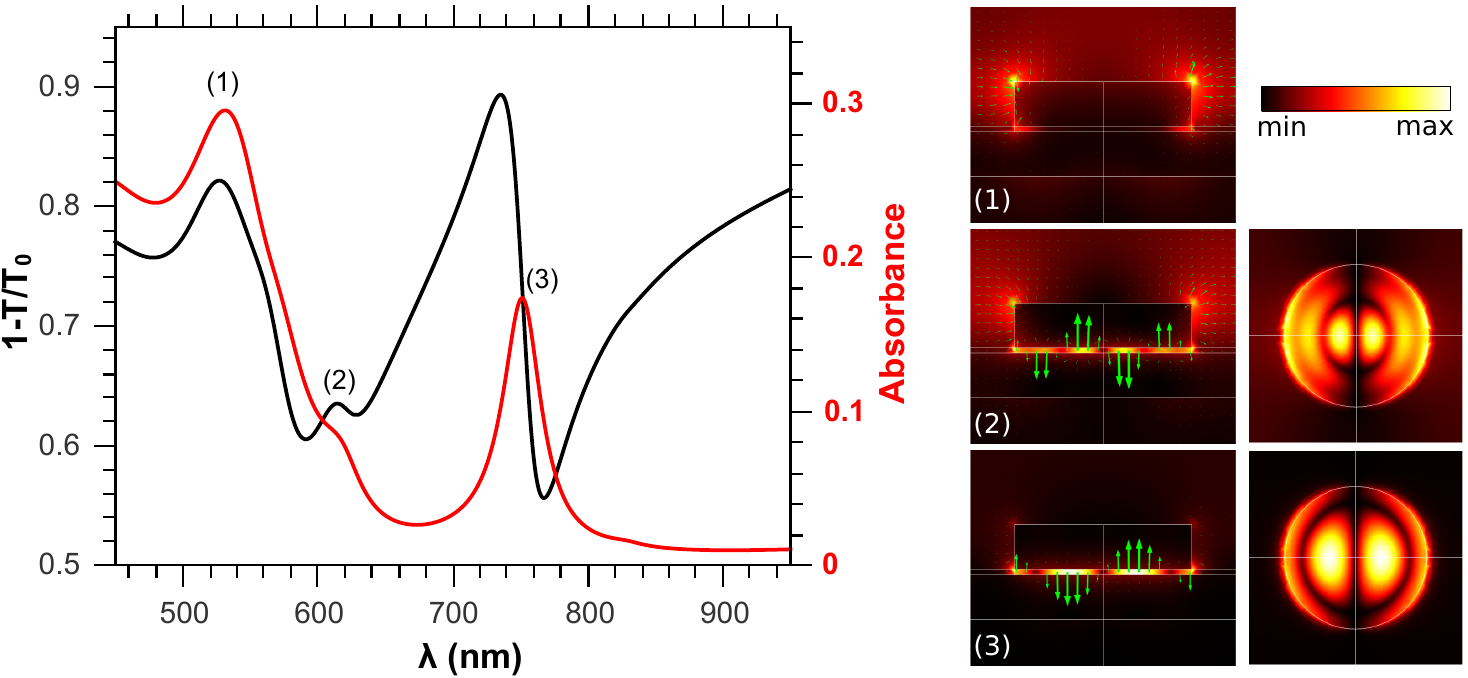}
	\caption{Extinction (black line) and absorbance (red line) spectra of the system depicted in Fig.~\ref{fgr:fig1}. The absorbance is computed inside the nanocylinder using $Q_e/(\epsilon_0 c E_0^2 a^2)$, where $Q_e$ is the power losses inside the nanocylinder and $E_0$ the amplitude of the incident electric field. The three field maps show the distribution of the complex electric field amplitude (color map) and the real part of the electric field (green arrows) for the three main resonances appearing in the spectrum.}
	\label{fgr:fig2}
\end{figure}

Figure \ref{fgr:fig2} shows the extinction spectrum $1-T/T_0$, where $T$ is the transmission through the grating (membrane+AuNCs) while $T_0$ is the transmission through the membrane alone (without the AuNCs), and the absorbance spectrum computed using $Q_e/(\epsilon_0 c E_0^2 a^2)$, where $Q_e$ is the power losses inside the nanocylinder and $E_0$ the amplitude of the incident electric field. For that simulation, a commercial finite element program has been used (Comsol). Three resonances appear, whose electric field distributions in the polarization $(Oxz)$ plane and in a $xy$-plane just under the bottom of the AuNCs are indicated on the right side of Fig.2. The short wavelength mode (labeled (1), $\lambda=525$ nm) corresponds to a dipolar mode localized on the top-edge of the nanocylinder, which has been observed both experimentally and theoretically in several systems made of flatten nanoparticles coupled to substrates \cite{maurer_enhancing_2013,maurer_coupling_2015,hohenau_plasmonic_2010}. Let us notice that very little light is confined in the spacer directly under the nanoparticle. 

This mode has been shown to be highly sensitive to refractive index changes is the superstrate, mostly due to the field enhancement at the top edge of the AuNCs. The two other modes ((2), $\lambda=615$ nm, and (3), $\lambda=750$ nm) are of very different nature: they correspond indeed to two so-called metal-insulator-metal localized-surface-plasmon (MIM-LSP) modes localized in the cavity in between the bottom of the nanoparticle and the top of the gold film. That cavity plays an important role in the formation of those modes, and will be called MIM-cavity in the following. As we will see in the following, those modes result from the formation of interference patterns in the MIM-cavity where propagating plasmons are excited by scattering of the incident plane wave.

\section{Analytical model}
Prior to any further numerical simulations, we present a simple analytical model which allows us to explain the formation of the MIM-LSP modes. Beside giving a good agreement of their resonance wavelengths and shape with the full numerical simulation, that model allows to simply assess the efficiency of the coupling between elastic and plasmonic modes.
\subsection{Model of the MIM-LSP}
The MIM-LSP modes supported by the investigated structure essentially result from the formation of resonant patterns within the circular MIM-cavity located in between the particle and the gold film, and limited transversally by the external circular edge of the bottom of the nanocylinder. Those resonant patterns originate in the modal superposition of propagative plasmons constrained to move under the bottom of the nanocylinder.
\begin{figure}[h!]
	\centering
	\includegraphics[width=8.5cm]{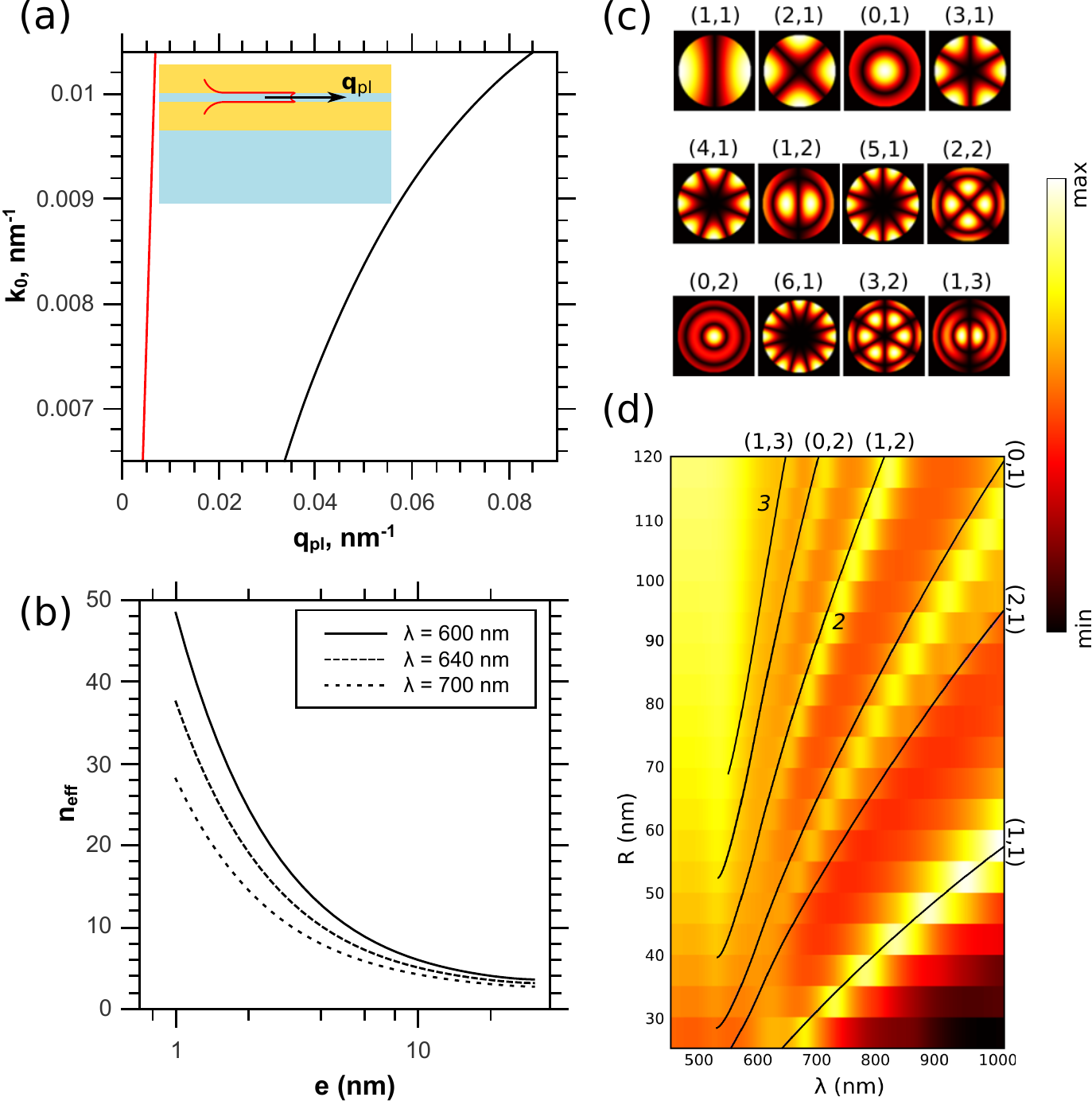}
	\caption{(a) black line: dispersion curve of the MIM-PSP supported by a structure made of air/Au(50nm)/SiO$_2$(6nm)/Au(50nm)/SiO$_2$(144nm)/air; red line: light line in silica ; (b) evolution of the effective index $n_{eff}=q_{pl}/k_0$ for three wavelengths as a function of the spacer thickness (logarithmic scale) ; (c) distribution of field for the twelve first modes as predicted by the analytical model ; (d) comparison between the resonance wavelengths for the MIM-LSP modes predicted by the analytical model and the full numerical simulation (Green's function method) of the absorption spectra computed as a function of the AuNC's radius for a tilted incidence angle ($\theta=\pi/3$).}
	\label{fgr:fig3}
\end{figure}

Let us suppose first that the particle has an infinite diameter: the system reduces then to an unidimensional system composed of air/Au(50nm)/SiO$_2$(6nm)/Au(50nm)/SiO$_2$(144nm)/air. It supports MIM propagative-surface-plasmon modes (MIM-PSP) whose wavevector $q_{pl}(\omega,e)$ depends both on the spacer thickness $e$ and the angular frequency $\omega=k_0 c=2\pi c/\lambda$, where $c$ is the speed of light in the vacuum. The corresponding dispersion curve is plotted in Fig.~3(a) for $e=6$nm. Notice that only the dispersion curve of the MIM-PSP has been plotted, but the system supports several other modes (guided modes inside the spacer and propagative plasmons on the top and bottom metal surface of the membrane). For that geometry, the effective index is $n_{eff}=q_{pl}/k_0\approx8.32$ for a wavelength in vacuum of $\lambda=2\pi/k_0=600$ nm, which corresponds to a mode wavelength of about $2\pi/q_{pl}=72$ nm along the interface. Hence the MIM-PSP mode has a typical length scale comparable to the nanocylinder diameter, which explains why those propagative waves can form resonant patterns on a surface with very sub-wavelength size. As indicated on Fig.~3(b), the effective index is strongly dependent of the spacer layer thickness, and increases, for a given wavelength, when the spacer is thinner.

From the numerical evaluation of $q_{pl}$ as a function of the frequency for a given thickness $e$, the wavelengths and the shapes of the MIM-LSP modes can be simply explained by describing them as a linear superposition of MIM-PSP propagating in different directions (given by the direction of $\mathbf{q}_{pl}$) under the nanoparticle. Such a superposition necessarily obeys the wave equation $\Delta_\parallel E_\alpha-q^2_{pl}E_\alpha=0$, where $\Delta_\parallel=\partial^2/\partial x^2+\partial^2/\partial y^2$ and $\alpha=x,y,z$. Let us notice that the wavevector $q_{pl}$ still depends both on the angular frequency of light $\omega$ and on the spacer thickness $e$, but those variables have been hidden for clarity. The solutions of that equation in polar coordinates $(\rho,\phi)$ are of the form $E_\alpha=A \cos(n\phi)J_n(q_{pl}\rho)$, where $A$ is an amplitude factor, and $J_n$ the Bessel function of first kind and of integer order $n$. Rigorously speaking, $A$ depends on $z$, but this dependency is very weak in the spacer layer (see profile of the mode in the inset of Fig.~\ref{fgr:fig3}(a)) and will be neglected in the following. In order to find the values taken by $q_{pl}$, and then the wavelengths $\lambda$, for each MIM-LSP mode, a proper boundary condition must be applied at the edge of the cavity, in $\rho=R$. As a simple rule, we notice that plasmonic modes generally lead to large intensity enhancement at sharp edges of metallic structures, which happens in particular in our case at the bottom circular edge of the AuNCs. Hence, the electric field being mostly vertical in the MIM-cavity as shown on the field distributions of Fig.~\ref{fgr:fig2} for modes (2) and (3), we choose values of $q_{pl}$ such that the $z$ component of the electric field reaches a maximum at $\rho=R$. This leads to $J'_n(q_{pl}R)=0$, where the prime denotes the derivative of the Bessel function with respect to its argument. We finally obtain the following expression for the $z$-component of the electric field:
$$E_{z}(\rho,\phi)=E_{0}\cos(n\phi)J_{n}(x'_{np}\rho/R)$$
with the dispersion equation:
\begin{equation}\label{Eq1}
\frac{1}{q_{pl}(\lambda,e)}=\frac{R}{x'_{np}}
\end{equation}
where $p$ is an integer larger than 1 and $x'_{np}$ are the zeros of the Bessel function's derivative. That relation is exactly the same as the one giving the cutoff wavelengths of the first TE modes in a circular metal waveguide. The first twelve   values of the coefficients $x'_{np}$ are: $x'_{1,1}=1.841$, $x'_{2,1}=3.054$, $x'_{0,1}=3.832$, $x'_{3,1}=4.201$, $x'_{4,1}=5.317$, $x'_{1,2}=5.331$, $x'_{5,1}=6.416$, $x'_{2,2}=6.706$, $x'_{0,2}=7.016$, $x'_{6,1}=7.501$, $x'_{3,2}=8.015$, and $x'_{1,3}=8.536$. The corresponding field distributions are shown in Fig.~\ref{fgr:fig3}(c). We can immediately notice on Fig.~\ref{fgr:fig2} that the two MIM-LSP modes labeled (2) and (3) correspond, owing to their field distribution, to the modes $(n,p)=(1,3)$ and $(n,p)=(1,2)$ in the analytical model. We compare in Fig.~\ref{fgr:fig3}(d) the position of the MIM-LSP modes predicted by the analytical model with the one resulting from the full numerical simulations of the absorption spectra as a function of the nanocylinder radius, for an incident plane wave at a tilted angle of 60$^o$, measured from the $(Oz)$ axis. It is necessary to use a tilted illumination in order to excite more MIM-LSP modes, as the normal incidence imposes a selection rule on the excited modes which must be symmetric with respect to the $(Oyz)$ plane and antisymmetric with respect to the $(Oxz)$ plane. In particular, the modes with even $n$ cannot be excited in normal incidence. The agreement between the simulation and the model is pretty good, slight red-shifts occur in the simulation compared to the model, more particularly for the $(n,p)=(0,1)$ mode. However, the variations of the wavelength with the radius of the particle match very well in both approaches.

\section{Phonon-plasmon coupling}
We will now assess the effect of a modification of the shape of the MIM-cavity under an elastic deformation. That deformation is physically induced by the movement of the AuNC above the spacer when a particular elastic mode is excited. Computer simulations of those modes are presented in the next section, however we show here that most of the physics is captured by considering the shape of the MIM-cavity independently of the AuNC.

We are interested here in two types of cavity deformations. In the first, called "radial breathing deformation", the cavity undergoes an isotropic oscillation of its radius at the elastic frequency, while the thickness of the cavity oscillates accordingly in phase opposition compared to the radius: when the cavity expands, it becomes thinner, and thicker when it contracts. In the second type of mode, called "multipolar deformation", the radius oscillates with a harmonic angle variation: $R(\phi)=R_0+\delta R\cos(m\phi)$, $m\ne 0$. In that case, we suppose that the top and bottom surfaces of the cavity keep the same area to the lowest order in $\delta R$, and that the thickness of the cavity stays mostly unchanged during the acoustic period.

\textit{Radial breathing deformation.} That deformation changes both $R$ and $e$. It is convenient to examine first a change of the radius of the MIM-cavity while keeping its thickness constant, and second a change of the thickness of the MIM-cavity while its radius is kept constant. Starting from $x'_{np}=R\,q_{pl}\left(\lambda,e\right)$ (Eq. \ref{Eq1}), we obtain:
$$
0=\frac{dR}{R}+\frac{dq_{pl}}{q_{pl}}=\frac{dR}{R}+\frac{1}{q_{pl}}\left[\left.\frac{\partial q_{pl}}{\partial \lambda}\right|_e d\lambda+\left.\frac{\partial q_{pl}}{\partial e}\right|_\lambda de\right]
$$
We introduce $S_{R}=\partial\lambda/\partial R|_e$, which represents the sensitivity of the MIM-LSP mode wavelength to a modification of the particle radius, keeping the thickness of the spacer layer constant. Writing $de=0$ in the previous expression, we find:
$$S_{R}=\left.\frac{\partial\lambda}{\partial R}\right|_{e}=-\frac{q_{pl}}{R}\frac{1}{\left.\frac{\partial q_{pl}}{\partial\lambda}\right|_{e}}$$
In a similar way, we obtain the sensitivity $S_{e}$ of the MIM-LSP mode wavelength to a modification of the spacer layer thickness:
$$S_{e}=\frac{e}{R}\left.\frac{\partial\lambda}{\partial e}\right|_{R}=-\frac{e}{R}\frac{\left.\frac{\partial q_{pl}}{\partial e}\right|_{\lambda}}{\left.\frac{\partial q_{pl}}{\partial\lambda}\right|_{e}}
$$
The factor $e/R$ allows to scale $S_e$ to values comparable to $S_R$, by taking similar deformation in thickness ($de/e$) and radius ($dR/R$) in both situations.

Next, we need to introduce the ratio $\nu_e$ of the radius deformation $dR/R$ compared to the thickness deformation $de/e$:
\begin{equation}\label{Eq2}
\nu_e=-\frac{dR}{R}/\frac{de}{e}
\end{equation}
The minus sign accounts for the phase opposition between the deformation in radius and the deformation in thickness. In case of a cylinder in vacuum, submitted to a force applied symmetrically on its top and bottom faces, this coefficient identifies with the Poisson's ratio $\nu$. However, in that situation $\nu_e$ cannot be predicted {\it a priori} but will be estimated later from the numerical simulations of the corresponding elastic eigenmode. As the resonance wavelength is a function of $e$ and $R$:
\begin{equation}\label{Eq3}
\lambda(e,R)\Rightarrow d\lambda=\left. \frac{\partial\lambda}{\partial e}\right|_R de +\left. \frac{\partial\lambda}{\partial R}\right|_e dR
\end{equation}
By combining equations \ref{Eq2} and \ref{Eq3}, we finally obtain the sensitivity $S_\nu$ of the MIM-LSP wavelength $\lambda$ under a breathing deformation:
\begin{equation}
S_{\nu}=\left.\frac{\partial\lambda}{\partial R}\right|_{\nu}=S_R-\frac{1}{\nu_e}S_e
\end{equation}

\textit{Multipolar deformation.} We suppose here that the thickness of the MIM-cavity does not change during the deformation, but that the radius of the MIM cavity is of the form $R(\phi)=R_0+\delta R\cos(m\phi)$ at the considered time of the acoustic deformation, $R_0$ being the radius of the particle at rest.We need to evaluate how much a MIM-LSP mode of order $(N,P)$ is modified under that deformation. For that purpose, we expand that mode on the basis of functions solution of the wave equation in cylindrical coordinates (see above):
$$E_z^{N,P}=\sum_{n=0}^\infty \alpha_{n}\cos(n\phi)J_n(q_{pl}\rho)$$
with $\alpha_{N}\approx 1$ and $\alpha_{n}<< 1,\;n\ne N$. The new boundary condition reads:
$$0=\left[E_z^{N,P}\right]'\left(R(\phi),\phi\right)=\sum_{n=0}^\infty \alpha_{n}\cos(n\phi)J_n\left[q_{pl}(R_0+\delta R \cos(m\phi))\right]$$
As the deformation is weak, the value of $q_{pl}R_0$ stays close to $x'_{NP}$, and we write it $q_{pl}R_0=x'_{NP}(1+\beta)$. To the first order in $\alpha_{n}$, $\beta$ and $\delta R/R_0$, with $J'_{N}(x'_{NP})=0$:
\begin{eqnarray*}
&0=\left\lbrace\cos(N\phi)\beta+\frac{\delta R}{2R_0} \left[\cos((m-N)\phi)+\cos((m+N)\phi)\right]\right\rbrace J''_N(x'_{NP})\\
&+\sum_{n\ne N}\alpha_{n}\cos(n\phi)J'_n(x'_{NP})
\end{eqnarray*}
We can verify that $m=0$ corresponds to the radial breathing deformation, and then $\alpha_{n}=0$ if $n\ne N$: the shape of the MIM-LSP mode is unchanged as the MIM-cavity is still exactly cylindrical. The only effect is a shift of the resonance wavelength. However, when $m\ne 0$, $\beta\ne 0$ only if $N=m-N$, which happens only if $m$ is even. In that case, the only mode affected by the deformation is $N=m/2$, and:
$$\beta=-\frac{\delta R}{2 R_0}x'_{NP}$$
$$\alpha_{N}=1$$
$$\alpha_{3N}=-\frac{\delta R}{2 R_0}\frac{J''_N(x'_{NP})}{J''_{3N}(x'_{NP})}$$
$$\alpha_{n}=0,\;n\ne N,3N$$
The wavelength of the deformed mode $N=m/2$ is then given by:
$$q_{pl}\left[\lambda,e\right]R_0=x'_{np}\left[1-\frac{\delta R}{2R_0}\right]\approx x'_{np}+R_0\left.\frac{\partial  q_{pl}}{\partial \lambda}\right|_e \delta\lambda$$
Finally, the sensitivity of the MIM-LSP mode wavelength under a multipolar deformation is independent on $m$ and reads:
$$S_{m}=\frac{\delta\lambda}{\delta R}=-\frac{q_{pl}}{2R}\frac{1}{\left.\frac{\partial q_{pl}}{\partial\lambda}\right|_{e}}=\frac{1}{2}S_R$$
Figure 4(a) shows the evolution of $-S_e$ and $S_R$ with the resonance wavelength $\lambda$ of the MIM-LSP mode, for $e=6$ nm. The sensitivity to the spacer thickness $S_e$ is negative, because when the spacer thickness increases ($\delta e>0$), the effective index $n_{eff}$ decreases: in order to satisfy the MIM-LSP excitation condition, the wavelength must then be shorter ($\delta \lambda<0$). However, $S_R$ is positive: an increase in diameter obviously does not change the effective index, and the wavelength must increase to compensate $\delta R>0$. We can additionally see that in the whole range of investigated wavelengths, $|S_e|<S_R$ by a factor of about 0.8 around $\lambda$=600 nm to about 0.5 around $\lambda=1$ $\mu$m.
\begin{figure}[h!]
	\centering
	\includegraphics[width=8.5cm]{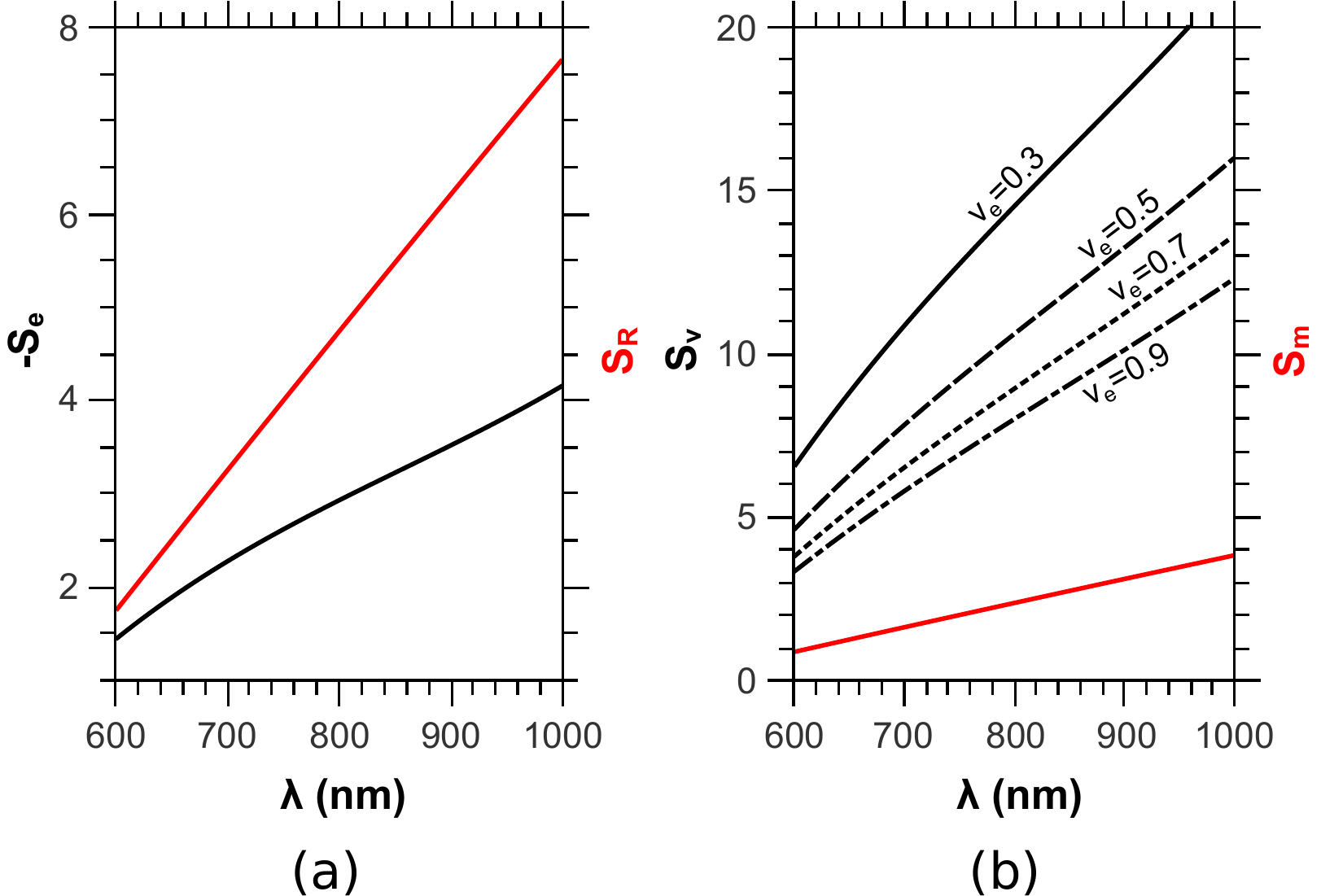}
	\caption{Evolution of the sensitivity of the MIM-LSP wavelength to: (a) black line (resp. red line): the thickness, (resp. radius) of the MIM-cavity at constant radius (resp. thickness); (b) a radial breathing deformation for different ratios $\nu_e$ (black lines) or a multipolar deformation (red).}
	\label{fgr:fig4}
\end{figure}
The most interesting result appears on Fig.~\ref{fgr:fig4}(b), which, similarly as Fig.~\ref{fgr:fig4}(a), shows the evolution of $S_\nu$ and $S_m$ with the wavelength. In particular, $S_\nu$ has been evaluated for several values of $\nu_e$ between 0.3 and 0.9. It appears clearly that $S_\nu$ is always larger that $S_m$, because the wavelength shifts $S_e$ and $S_R$ cumulate when the cavity undergoes a radial breathing deformation, as an increase in radius generally induces a decrease in thickness. For example, $S_\nu$ and $S_m$, which are both positive, differ for $\nu_e=0.5$ by a factor of about 5.5 to 4.5 from 600 nm to 1 $\mu$m, the change in wavelength induced by a radial breathing deformation being then significantly larger that the one induced by a multipolar deformation which does not change the MIM-cavity thickness.

\section{Full numerical simulations}
For the numerical simulations, a commercial finite element method has been used (Comsol) in order to be able to simultaneously investigate the plasmonic, phononic, and coupled acousto-plasmonic aspects. The radio-frequency module is employed for the photonic simulations, while the structural-mechanics module is used for the phononic simulations.
\subsection{Phononic properties}
First, we investigate the elastic properties of the structure described in Fig.~\ref{fgr:fig1} ($R=100$ nm, $h_p=50$ nm, $e=6$ nm, $h_f=50$ nm, $H=144$ nm, $a=300$ nm) by computing the elastic eigenmodes and eigenfrequencies of a unit cell of the grating. Free interfaces boundary conditions are applied on the topmost surfaces of the AuNC and the silica spacer, and on the bottom surface of the 144nm-thick silica layer, while periodic boundary conditions are applied on the lateral boundary of the unit cell. The elastic parameters for silica and gold have been taken in Royer and Dieulesaint \cite{royer_elastic_2000}. Gold is anisotropic (face centered cubic), and the crystal is oriented such as $xy$-planes correspond to $(001)$ crystallographic directions of gold; silica is isotropic.
\begin{figure}[h!]
	\centering
	\includegraphics[width=8.5cm]{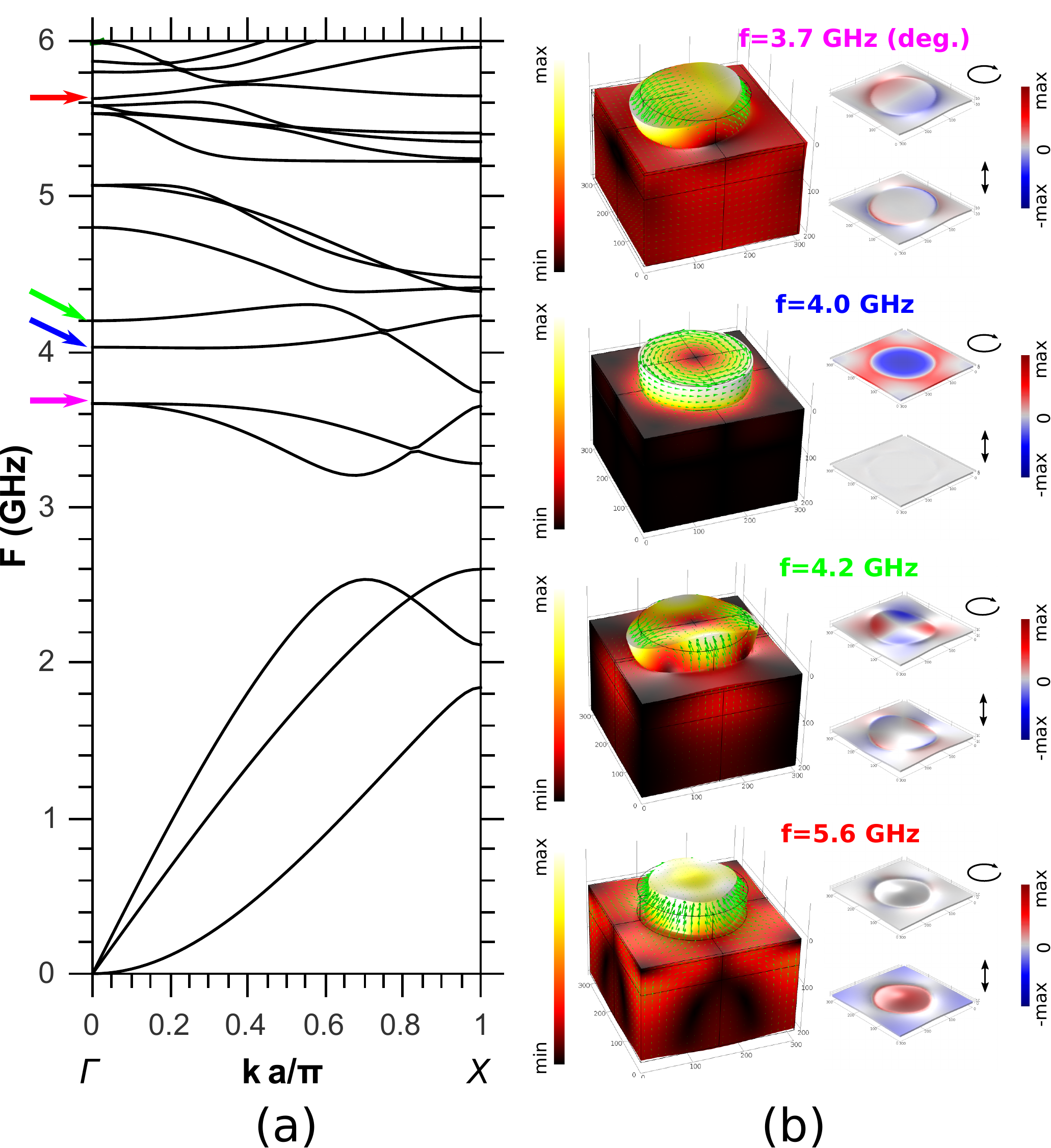}
	\caption{(a) Band diagram along the $\Gamma X$ direction ; (b) For each of the four frequencies indicated by an arrow in (a) at the $\Gamma$ point: on the left side is plotted the displacement ($\mathbf{u}=(u_x,u_y,u_z)$) maps in 3d (color map: $|\mathbf{u}|$, green arrows: $\Re(\mathbf{u})$), while on the right side is plotted the deformed representation of the silica spacer under the AuNC, where color maps show the $z$ component of $\nabla \times \mathbf{u}$ ($\circlearrowright$, top) and $\partial_z u_z$ ($\updownarrow$, bottom).}
	\label{fgr:fig5}
\end{figure}

Figure 5(a) shows the resulting band diagram along the $\Gamma X$ direction, where a partial band-gap is opened around 3GHz. Overall, the structure supports a lot of modes, most of which are strongly dispersive and have elastic energy spread all around the AuNCs. However, for five modes at the $\Gamma$ point most of the energy is localized in or close to the AuNC, as indicated by their displacement distribution. Their respective frequencies are 3.7 Ghz (two degenerated modes along the $x$ and $y$ directions), 4.0 GHz, 4.2 GHz, and 5.6 GHz, their displacement maps are indicated in Fig. 5(b). The color map shows the amplitude of the displacement vector $\mathbf{u}=(u_x,u_y,u_z)$ (real at the $\Gamma$ point), while the green arrows indicate the displacement itself. The deformation is obviously exaggerated in order to more clearly show the shape of the mode. On the right side of each displacement maps is represented the deformed 6-nm-thick spacer layer. The deformation is again amplified for clarity, and the colorscale represents either the $z$ component of $\nabla \times \mathbf{u}$ ($\circlearrowright$, top) or $\partial_z u_z$ ($\updownarrow$, bottom). The first quantity indicates the rotation experienced by the MIM-cavity around the vertical ($z$) axis, while the second shows its relative change in thickness. These maps emphasize the differences between the three excited modes and their potential effect on the MIM-LSP modes.

The two lowest frequency modes at 3.7 GHz are degenerated, and correspond to flexural deformations along the $(Ox)$ or the $(Oy)$ axis of the square grating. As a result, the nanocylinder and the MIM-cavity oscillate together around either the $(Ox)$ or the $(Oy)$ axis, and very small deformation occurs in the MIM-cavity: its shape is mostly unchanged and the coupling between that elastic mode and the MIM-LSP is expected to be weak. Yet, we will have to differentiate the $x$ and $y$ flexural modes in the plasmon-phonon coupling as the incident electric field is polarized along the $x$-axis. 

The next mode at 4.0GHz is an azimuthal shear deformation mode where the displacement is essentially orthoradial, enhanced close to the circular top edge of the nanocylinder but much lower at its bottom. Under that deformation, the MIM-cavity undergoes no compression (constant thickness and radius) but a clear rotation around the vertical axis. That displacement is coupled to a very weak radial breathing movement which corresponds to a small increase of the volume of the particle, which reaches a maximum every half an acoustic period. Again, this mode is not expected to strongly couple with the MIM-LSPs. 

The mode at 4.2 GHz is a quadrupolar mode, enhanced on the top part of the nanoparticle, with displacement maxima along the main directions of the grating. That mode is symmetric compared to both $Oxz$ and $Oyz$ planes, and does not strongly affect the volume of the nanoparticle, as the expansion/contraction along the $x$ direction is balanced by the opposite displacement in the $y$ direction. The MIM-cavity undergoes a quadrupolar deformation, which results in a slight vertical compression / dilatation at its circular edge: the thickness of the MIM-cavity is hardly modified except close to its border. Hence, the radius of the cavity follows at that frequency an azimuthal-angle dependency of the form $R(\phi)=R_0+\delta R\cos(m\phi)$ with $m=2$.

Finally, the high-frequency mode at 5.6 GHz clearly shows a vertical compression / dilatation of the nanocylinder, while its average radius correspondingly increases and decreases during an acoustic period. The displacement map is mostly symmetric compared to any plane perpendicular to the substrate containing the particle revolution axis (despite slightly deformed by the closest neighboring particles along the $Ox$ and $Oy$ axes). Similarly, the MIM-cavity keeps its circular shape during the deformation, its radius alternately increases and decreases while its thickness decreases and increases. However, contrary to the 4.0GHz mode, the rotation of that mode around the perpendicular axis is very small. Finally, the MIM-cavity at that frequency clearly undergoes a radial breathing deformation as presented in the analytical model.

\subsection{Phonon-plasmon interaction}
In that section, we investigate the influence of the five previous elastic modes on the optical properties of the system. For that purpose, the deformed geometry corresponding to each mode is computed for different phases $\psi=\Omega t=2\pi F t$ during half an acoustic period, where $F$ is the eigenfrequency of the considered elastic mode. The origin of phase is chosen such as $\psi=0$ corresponds to the structure at rest. The amplitude of the deformation is chosen such as the maximum of the elastic displacement in the nanocylinder is taken equal to 2\% of its radius, which gives here $u_{max}=2$nm. Then, that deformed structure is used to compute the extinction spectrum under the same illumination conditions as in section 2, at the given phase $\psi$. Let us mention that, at the scale of the crystalline network, the corresponding deformation is of about $u_{max}/R \approx 2$\%, which is huge and not physically meaningful. For instance, in reference \cite{soavi_ultrasensitive_2016}, authors estimate that the length increase of their 150-nm-long nanorods is on the order of 5pm for the considered acoustic modes, which gives a deformation of about 3.3 10$^{-5}$. For numerical purpose it is necessary to use larger values, however the obtained quantities can then be scaled to smaller deformation in order to obtain realistic figures. In the following, the modulated spectra will be analyzed to assess the strength of the coupling, using two methods. In the first, the shift in wavelength of the different plasmon modes is evaluated as a function of the acoustic phase $\psi=\Omega t$, and in the second, the relative variation of the transmission is computed for realistic deformations, similarly to what can be realized in a typical pump-probe experiment.
\begin{figure}[h!]
	\centering
	\includegraphics[width=8.5cm]{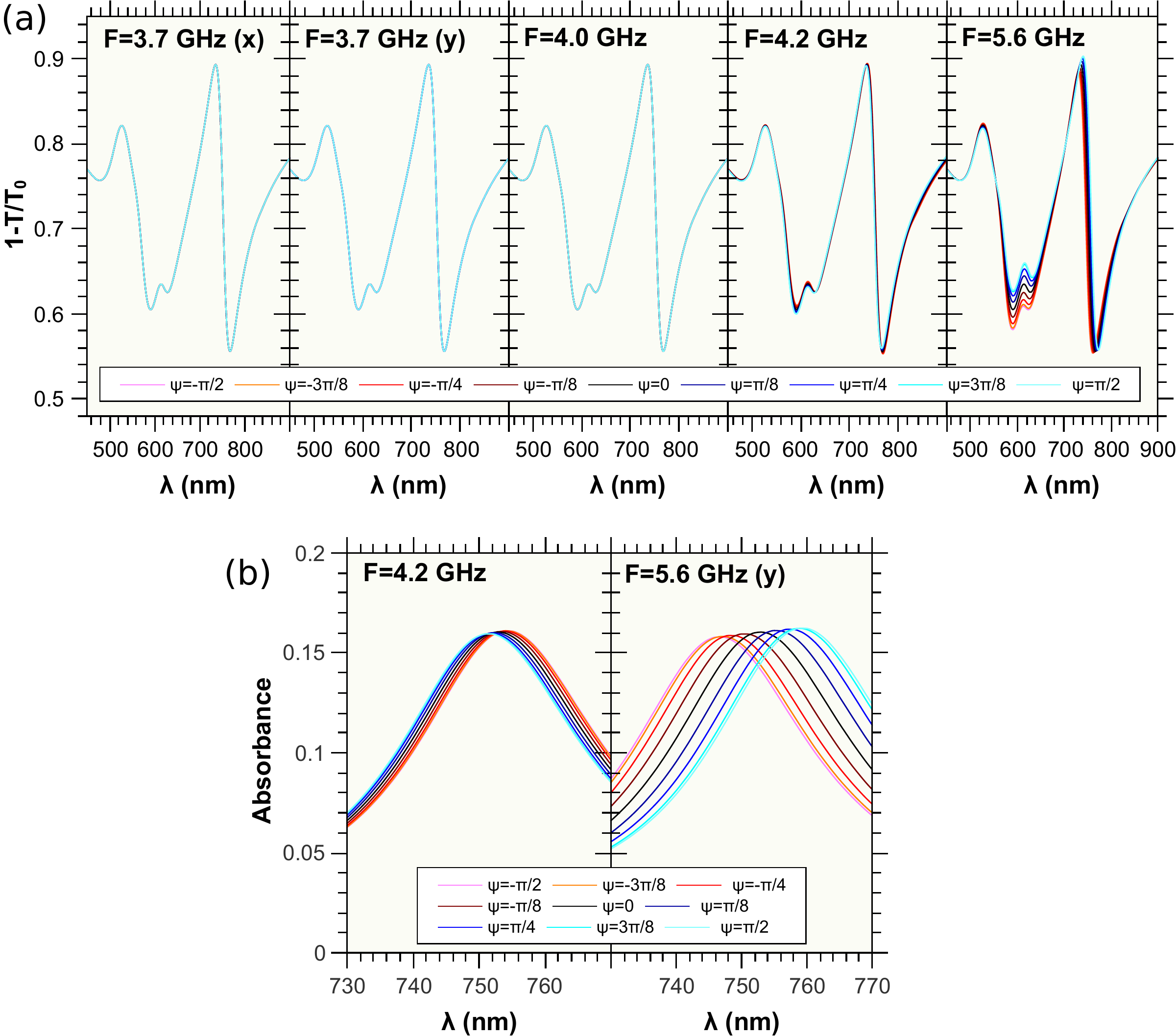}
	\caption{(a) Evolution of the extinction spectra $1-T/T_0$ for different values of the elastic phase for the five modes of Fig.~\ref{fgr:fig5} (T: transmission through the membrane+AuNCs, $T_0$: transmission through the membrane without the AuNCs); (b) Evolution of the absorbance spectra for different values of the elastic phase for the two elastic modes at 4.2GHz and 5.6GHz, around the wavelength of the main $(n,p)=(1,2)$ MIM-LSP mode.}
	\label{fgr:fig6}
\end{figure}

First, we present on Fig.~\ref{fgr:fig6}(a) the modulated extinction spectra for the five elastic modes chosen in the previous section. A closer view of the evolution of the absorbance spectrum around the wavelength of the MIM-LSP mode at 750 nm is shown in Fig.~6(b) for the two highest-frequency elastic modes. The differences between the five modes are very obvious. First, the two flexural modes at 3.7GHz and the azimuthal shear deformation mode at 4.0GHz produce no changes neither in the extinction nor in the absorption spectra. That result is obvious for the 4.0GHz mode as the displacement is essentially orthoradial and then does barely move or deform any interfaces in the system. For the flexural modes, despite the fact that interfaces move, neither the shape of the particle nor the shape of the MIM-cavity is modified, as they both rotate undeformed around the $x$- or the $y$- axis. The second mode at 4.2GHz however induces noticeable modifications which are more easily observable on the absorption spectrum around the main resonance at 750nm  (MIM-LSP mode $(n,p)=(1,2)$). Finally, the high frequency mode at 5.6GHz produces the largest modifications, easily observable both in the extinction and the absorption spectra, and mostly in the range of wavelengths where the MIM-LSP modes are excited. In the following, we focus only on the quadrupolar mode at 4.2GHz and the vertical breathing mode at 5.6GHz.

\begin{figure}[h!]
	\centering
	\includegraphics[width=8.5cm]{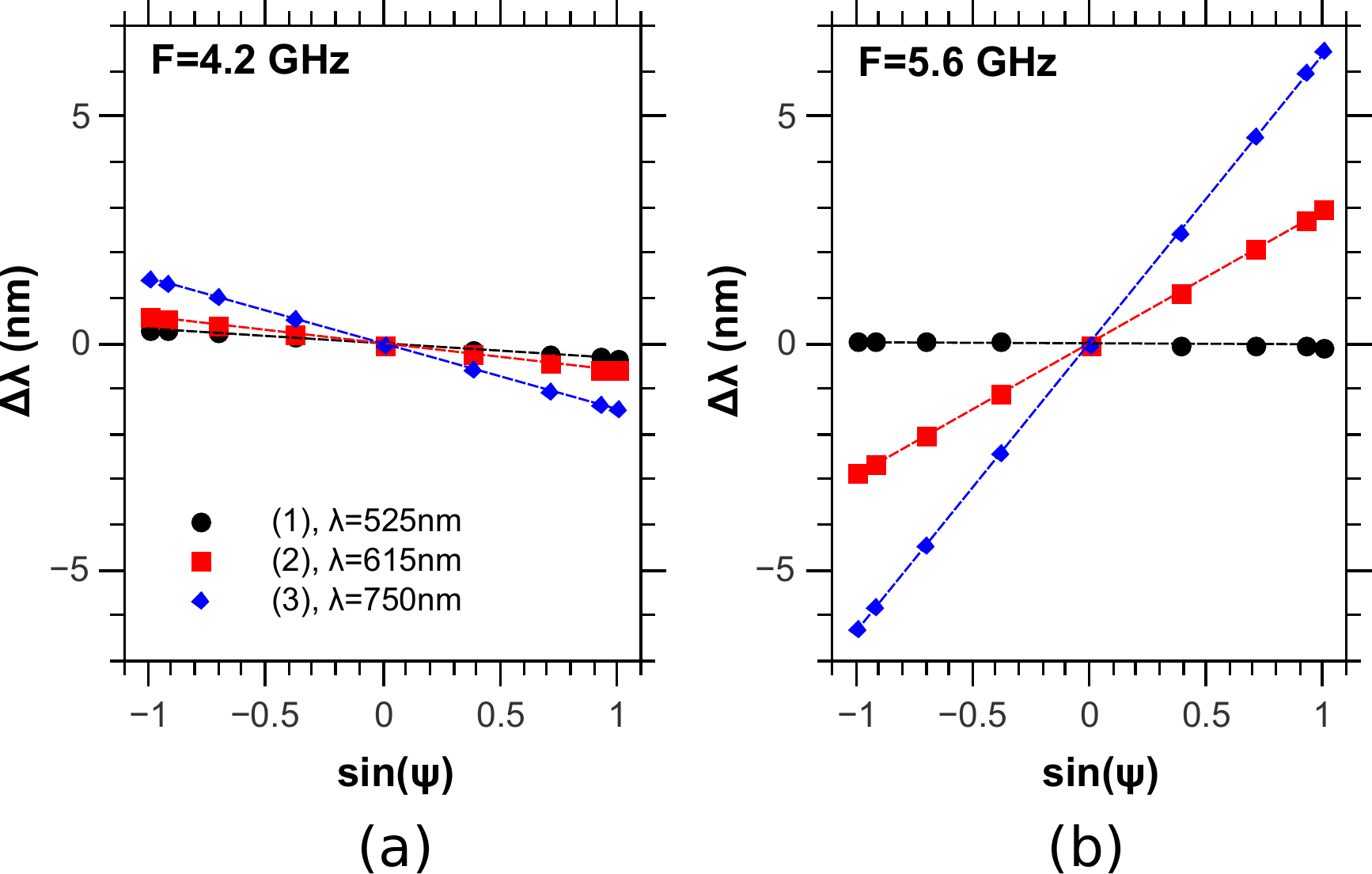}
	\caption{Modulation of the wavelength of the three plasmon modes as a function of the sine of the acoustic phase of the two elastic modes: (a) F=4.2 GHz ; (b) F=5.6 GHz. Symbols are the wavelengths obtained directly from absorbance spectra while dashed lines are linear fits.}
	\label{fgr:fig7}
\end{figure}

Figure \ref{fgr:fig7} shows the evolution of the variation of the resonance-wavelength of the three localized plasmons modes (the short-wavelength, top-face-localized-dipole at $\lambda$ = 525 nm, and the two MIM-LSP modes at 615 nm ($(n,p)=(1,3)$) and 750 nm ($(n,p)=(1,2)$)) as a function of the sine of the phase of the elastic wave when coupled to the quadrupolar mode (Fig.~\ref{fgr:fig7}(a)) and to the vertical breathing mode (Fig.~\ref{fgr:fig7}(b)). The position of the different peaks have been determined by fitting the absorbance spectra with Lorentzian functions. It appears clearly that the wavelengths depend linearly on the sine of the phase. Very weak deviations from that linear law might be noticeable, which must be attributed to the large value of the maximum displacement (2 nm). A linear fit of $\Delta \lambda$ as a function of $\sin(\psi)$ is plotted as dashed lines in Fig.~\ref{fgr:fig7}. For the elastic mode at $F$=4.2 GHz, the MIM-LSPs have, for the imposed value of 2 nm maximum displacement, a linear shift of $\delta\lambda/\sin(\psi)$=-0.6 nm (resp. $\delta\lambda/\sin(\psi)$=-1.5 nm) for the 615 nm (resp. 750 nm) MIM-LSP modes. The shortest-wavelength mode at 525 nm is less sensitive with $\delta\lambda/\sin(\psi)$=-0.3 nm. The wavelength shifts obtained with the $F$=5.6 GHz elastic mode are much larger for the two MIM-LSP modes with $\delta\lambda/\sin(\psi)$=2.9 nm for the 615 nm mode and $\delta\lambda/\sin(\psi)$=6.4 nm for the 750 nm mode. However, the 525 nm mode shows almost no wavelength dependency with a $\delta\lambda/\sin(\psi)$=-0.02 nm. Hence, the wavelength modulation amplitude is clearly enhanced whith the F=5.6GHz mode compared to the 4.2GHz, by a factor of 4.9 for the $\lambda=615$ nm MIM-LSP and by a factor of 4.4 for the $\lambda=750$ nm MIM-LSP.

Those extracted values of $\delta\lambda/\sin(\psi)$ have been obtained for a maximum displacement of $u_{max}=2$ nm as explained earlier. In order to compare those numerical results to the figures obtained with the analytical model, we need to evaluate the quantities $\delta\lambda/\delta R$ where $\delta R$ is the radius variation of the MIM-cavity for each of the two considered elastic modes. By a direct evaluation on the simulated displacement maps, those values have been estimated to $\delta R=0.59$ nm for the quadrupolar mode and $\delta R=1.36$ nm for the vertical breathing mode. For the $F=5.6$ GHz elastic mode, we need as well an estimate of the deformation ratio $\nu_e$. We find that $de/e\approx 0.015$ for $u_{max}=2$ nm, which gives $\nu_e\approx 0.92$. The comparison between the sensitivities obtained in the numerical and the analytical approach are indicated in Table \ref{tb1}.
\begin{table}
	\caption{Summary, for the three localized plasmon modes, and for the two elastic modes at 4.2GHz and 5.6GHz, of the wavelength shifts per $\sin(\psi)$, $\delta\lambda/\sin(\psi)$, or of the wavelength shift per maximal variation of the cavity's radius $\delta\lambda/\delta R$ (mode at 525 nm is not concerned (n.c.)). Simulated (s) and analytical (a) values are compared.}
	\centering
	\begin{tabular}{|c|ccc|ccc|}
	\hline
	& & F=4.2GHz & & & F=5.6GHz &\\
	\hline
	$\lambda_{LSP}$ &$\frac{\delta\lambda}{\sin(\psi)}$ & $\frac{\delta\lambda}{\delta R},s$&$\frac{\delta\lambda}{\delta R},a$&$\frac{\delta\lambda}{\sin(\psi)}$ & $\frac{\delta\lambda}{\delta R},s$&$\frac{\delta\lambda}{\delta R},a$\\
	525 nm & -0.3 & n.c. & n.c. & 0.02& n.c. & n.c. \\
	615 nm & 0.59 & 1. & 0.93 & 2.92 & 2.15 & 3.53 \\
	750 nm & 1.45 & 2.46 & 1.9 & 6.37 & 4.68 & 6.90 \\
	\hline
    \end{tabular}\label{tb1}
\end{table}

The agreement is overall qualitatively correct in the four cases, with the largest discrepancy for the 615 nm LSP/5.6GHz elastic mode, and the best agreement for the 615 nm MIM-LSP/4.2GHz elastic mode. The sensitivity is underestimated by the analytical model for the 4.2GHz mode but overestimated for the 5.6GHz mode. Differences might partly be attributed to the fact that the analytical model is very simplified as it does not take into account the precise shape of the cavity, nor the fact that the MIM-PSP wave vector is complex. However, the strong increase in sensitivity with the radial breathing elastic mode compared to the quadrupolar mode is correctly obtained in both cases, which makes us think that the physics of the coupling is correctly captured even with that simplified description. In particular, the enhanced wavelength shift with the high frequency acoustic deformation is clearly due to the cumulative effect of the expansion of the MIM-cavity (radius) and the contraction of the spacer layer (thickness).

Finally, we estimate in that section the time-modulation of the transmission spectra for more realistic deformations of the structure under coupling with the quadrupolar and radial breathing acoustic deformations. 
\begin{figure}[h!]
	\centering
	\includegraphics[width=8.5cm]{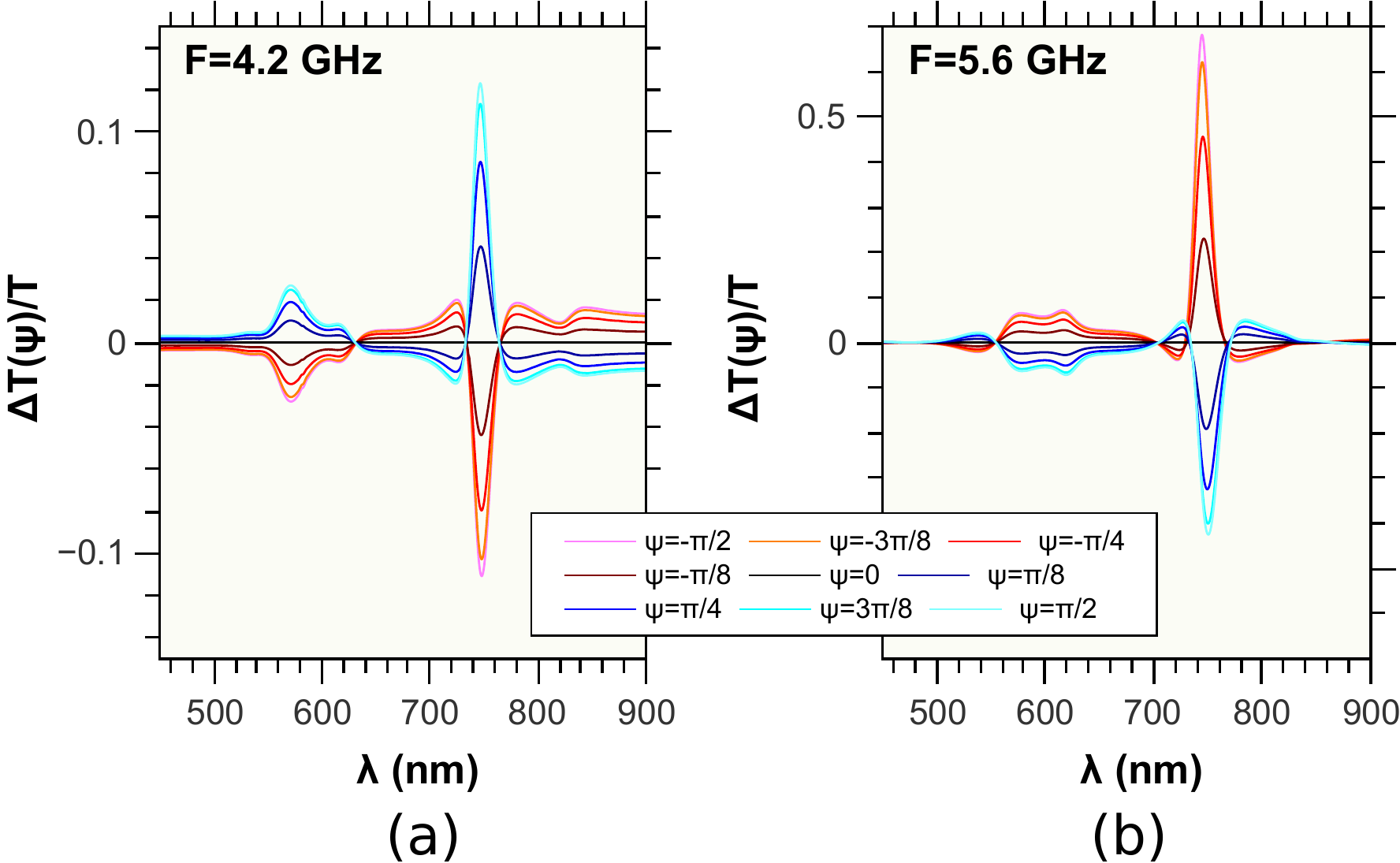}
	\caption{$\Delta T /T=(T^*-T)/T$ for the different phases $\psi$ of the two elastic modes at F=4.2GHz, (a), and F=5.6GHz, (b), where $T^*(\lambda)$ is the transmission through the deformed structure at phase $\psi$ and $T(\lambda)$ the transmission through the structure at rest.}
	\label{fgr:fig8}
\end{figure}
For that, we start from the fact that in the limit of weak deformation, the evolution of the transmission spectrum $T^*(\lambda)$ is a harmonic function of time for every fixed wavelength. This is relied here to the fact that the displacement vector $\mathbf{u}$ is real: at a certain time $t$ of the acoustic period, the displacement will then be at every point $\mathbf{u} \sin(\Omega t)$: changing the amplitude of the displacement is completely equivalent to changing the sine of the phase $\psi=\Omega t$ of the acoustic mode. Despite the fact that the transmission might oscillate with $2\Omega$ in some situations \cite{rolland_acousto-optic_2012}, this is not the case here. Figure \ref{fgr:fig8}(a) shows, for the two acoustic modes, the relative difference $\Delta T/T=(T^*(\lambda)-T(\lambda))/T(\lambda)$ between the modulated $T^*(\lambda)$ and the unmodulated transmission $T(\lambda)$ as a function of the acoustic phase $\psi$: $\Delta T/T$  is a mostly symmetric function of the phase, and whatever the wavelength, the modulation of the optical transmission has the same period than the elastic mode. Following that idea, we compute for every wavelength the slope $a(\lambda)$ of the tangent to the modulated transmission $T^*(\lambda)$ in $\psi=0$, from which we can evaluate the relative variation of transmission $\Delta T/T=(T^*(\lambda)-T(\lambda))/T(\lambda)$ for any arbitrarily small deformation of the structure using:
$$\frac{\Delta T}{T}(t)=\frac{B}{B_0}\, a(\lambda) \sin(\Omega t)$$
where $B$ is a coefficient corresponding to the chosen “realistic” deformation $B=u_{max}/R$, and $B_0=0.02$ is the reference deformation used in the simulation. Figure 9 shows the relative variation in transmission of the modulated system under the two elastic deformations as a function of time. In case of a pump-probe experiment, the amplitude of the modulation should be damped due to the losses, which are not taken into account here \cite{hartland_optical_2011}. The two diagrams have been plotted using a typical deformation of the structure of $B=u_{max}/R=3\,10^{-5}$, which correspond to a maximum displacement of about 3 pm inside the AuNC. For both acoustic modes, the obtained numbers are comparable to the values obtained experimentally in \cite{soavi_ultrasensitive_2016}, of about 10$^{-4}$, and reaches 8.9 10$^{-4}$ close to the 750 nm MIM-LSP mode when the structure is deformed by the radial breathing elastic mode: this corresponds to an increase of 4.6 in the oscillation amplitude of $\Delta T$ compared to the amplitude obtained with the quadrupolar elastic mode.
\begin{figure}[h!]
	\centering
	\includegraphics[width=8.5cm]{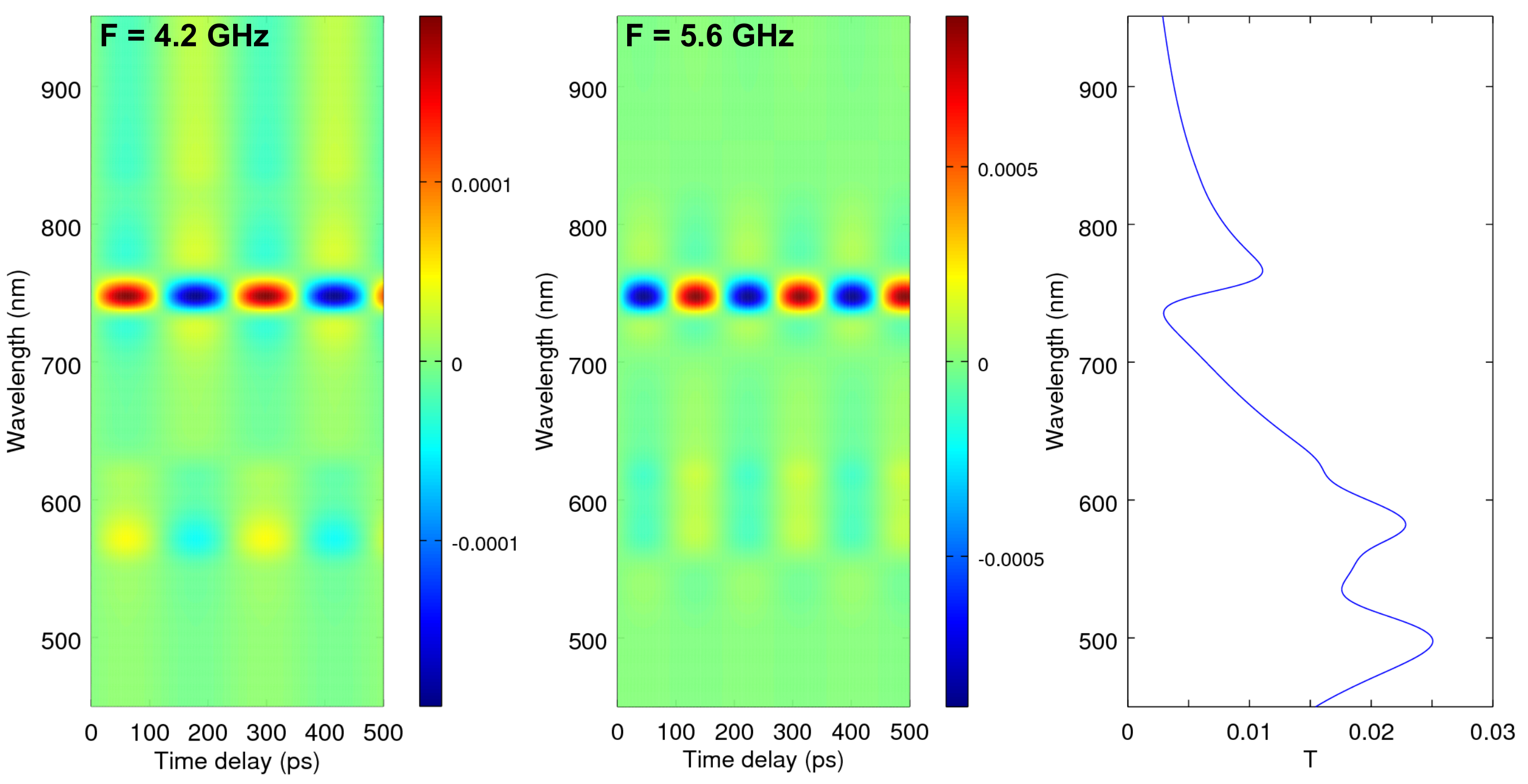}
	\caption{Time evolution of the relative variation of the transmission (T*-T)/T compared to the nanostructure at rest under the two elastic modes at F=4.2 GHz and F=5.6 GHz, computed with a deformation of $A=u_{max}/R=3.10^{-5}$. The blue curve is the unmodulated transmission $T(\lambda)$.}
	\label{fgr:fig9}
\end{figure}

\section{Conclusion}
Using a combination of finite elements numerical simulations and a simplified analytical model, we have provided physical insight in both the origin of the MIM-LSP modes supported by AuNCs on a thin silica film coating a metal interface, and their coupling with elastic modes supported by the same system, for which the movement of the AuNCs induces a deformation of the cavity formed by the portion of the silica spacer underneath. Due to the nature of the MIM-LSP, the three lowest-frequency elastic modes (flexural and azimuthal shear deformation modes), which do not modify the shape of that cavity, do not change significantly the optical response of the system. However, stronger effect is obtained when the AuNC, and then the MIM-cavity, experiences a quadrupolar deformation or even better a radial breathing movement. The latter clearly results in an enhanced wavelength shift and transmission modulation, due to the cumulative effect of the increase/decrease of the cavity's radius and the corresponding contraction/dilatation of its thickness. Compared to the quadrupolar deformation which only affects the radius, both the wavelength shifts and the transmission modulation are increased by a factor of about 4.6. That structure could give interesting results in a typical pump-probe experiment, despite the fact that efficiency of the coupling between the probe and the vibration of the particle is not precisely known, but only estimated from other publications. We believe that such a study can be of interest for the fundamental understanding of the coupling mechanisms between localized plasmon modes and elastic modes, and to research groups working on fabrication nano-opto-mechanical devices.

\paragraph {\it Acknowledgement}
This work was partially supported by VisionAIRR project "PolarEP" of R\'egion Nord-Pas-de Calais (France).

\end{document}